\documentclass[conference]{IEEEtran}

\makeatletter
\let\old@ps@headings\ps@headings
\let\old@ps@IEEEtitlepagestyle\ps@IEEEtitlepagestyle
\def\confheader#1{%
	\def\ps@headings{%
		\old@ps@headings%
		\def\@oddhead{\strut\hfill#1\hfill\strut}%
		\def\@evenhead{\strut\hfill#1\hfill\strut}%
	}%
	\def\ps@IEEEtitlepagestyle{%
		\old@ps@IEEEtitlepagestyle%
		\def\@oddhead{\strut\hfill#1\hfill\strut}%
		\def\@evenhead{\strut\hfill#1\hfill\strut}%
	}%
	\ps@headings%
}
\makeatother

\ifCLASSINFOpdf
  
\else
 
\fi

\ifCLASSOPTIONcompsoc
\usepackage[caption=false, font=normalsize, labelfont=sf, textfont=sf]{subfig}
\else
\usepackage[caption=false, font=normalsize]{subfig}
\fi
\usepackage{lipsum}%
\usepackage[dvipsnames]{xcolor}

\usepackage{balance}
\usepackage{multicol}   
\usepackage{cite}
\usepackage{gensymb}
\usepackage{multirow}
\usepackage{graphics}
\usepackage{epsfig}
\usepackage{graphicx}
\usepackage{epstopdf}
\usepackage{textcomp}
\usepackage{amsmath}
\usepackage{mathtools}
\usepackage{filecontents}
\usepackage{lipsum,color}
\usepackage{amssymb}
\usepackage{float}
\usepackage{algorithm,algorithmic}

\usepackage{times} 
\usepackage{amsthm}  

\usepackage{amsfonts}

\begin{document}
\title{Achieved Throughput of Hovering UAV-Based Optical Wireless Communications}
\author{
	Mohammad Taghi~Dabiri, Seyed Mohammad Sajad~Sadough~and~Himan~Savojbolaghchi\\ 
	{\it Department of Electrical Engineering, Shahid Beheshti University G. C., 1983969411 Tehran, Iran}\\ 
	Email: m\_dabiri@sbu.ac.ir, s\_sadough@sbu.ac.ir, h.savojbolaghchi@mail.sbu.ac.ir
}

\maketitle

\begin{abstract}
Recently, free-space optical (FSO) communication systems have obtained a wide range of applications for communications between unmanned aerial vehicles (UAVs) because they offer several advantages with respect to the traditional RF communication systems.
%
To investigate the performance of UAV-based FSO systems in terms of achievable data rates, we derive the exact closed-form expression for the mean achievable rates of UAV-to-UAV transmission. Our results may serve to investigate the interaction between different FSO system parameters and the average data rates achieved by UAV-based optical wireless communications.
\end{abstract}
\IEEEpeerreviewmaketitle
\section{Introduction}

Unmanned aerial vehicle (UAVs) (commonly known as drone) is considered to generate of next wireless networks \cite{9200666,dabiri2020analytical}.
Recently, the employment of UAVs for establishing non line-of-sight (NLOS) free-space optical (FSO) communications has considered for increasing the flexibility of FSO transmission and creating high throughput wireless connectivity \cite{alzenad2018fso,fawaz2018uav,8882271,dong2018edge,ramdhan2017tree,dabiri2019blind,hs2019spatial, kaadan2014multielement, kaadan2016spherical, najafi2017statistical,8827664, 9257013,mypaper}.
The related literature survey is as follows. In \cite{alzenad2018fso}, the authors investigated the feasibility of a backhaul link where the connectivity between the core network and access is based on FSO communication links.
In \cite{fawaz2018uav}, the integration of UAVs as  mobile relays into the conventional FSO relay systems has been introduced to enhance the performance of the existing FSO relay systems.
%
The effect of physical parameters such as position 	and height of obstacles is studied in \cite{8882271} to find the optimal location of a hovering UAV  acting as optical relay.
In \cite{ramdhan2017tree}, the authors proposed an optical UAV-assisted network    that collects data from sensors based on multiple BS     and multiple collection trees.
In \cite{dabiri2019blind}, the authors proposed a novel technique to detect optical signals for fast moving platforms where the sampling time is suffering from random errors.
Spatial tracking of optical signal under UAV fluctuations is studied in \cite{hs2019spatial} for the receiver equipped by photodetector array.


Channel modeling which is a critical issue for practical deployment of FSO systems, has been addressed in \cite{kaadan2014multielement, kaadan2016spherical, najafi2017statistical,mypaper}, for instance.
In \cite{kaadan2014multielement}, a new approach to alignment of a mobile FSO link was proposed.
The new analytical characteristics for a UAV-based FSO link were proposed in \cite{kaadan2016spherical} to study the intersection surface of a spherical cap.
The geometrical loss of a UAV-assisted FSO link is analyzed in \cite{najafi2017statistical} wherein several UAVs hover above an area and acting as mobile access nodes for a large number of mobile users. 
However, the effect of atmospheric turbulence degradation is not considered in \cite{kaadan2014multielement, kaadan2016spherical,najafi2017statistical} which can be significantly degrades the performance of considered FSO link.
More recently, by considering the joint effect of position and AoA fluctuations along with atmospheric turbulence, an accurate channel models for UAV-based FSO links were proposed in \cite{8827664,mypaper}.

According to \cite{mypaper}, the UAV channel model is characterized by different random tunable parameters such as optical beam width and the receiver FOV. More precisely, it is not straightforward to tune these parameters so as to maximize the average throughput of such systems over a wide range of FSO channel conditions. 
It is hence of practical interest to find the optimum values for tunable FSO link parameters in order to enhance the capacity of the considered UAV-assisted FSO links. 
Please note that finding the optimal values for the aforementioned channel parameters by performing Monte Carlo simulations requires a long processing time in FSO systems.
However, Monte Carlo analysis is used in this work to confirm the accuracy of the derived analytic expressions. In this work, we propose the closed-form expressions for average throughput of UAV-to-UAV  transmission system in terms of well-known elementary functions. 
Our results can be used in order to analysis of the trade-off between the required capacity and main tunable system parameters over a wide range of turbulence channel conditions.    
\begin{figure*}
\normalsize
\begin{align}
\label{uu}
f_{h}^{uu}(h) =&~ \sum_{m=0}^M \frac{\mathcal{H}(m) m!}{2}\delta(h) + C_a h^{\gamma^2-1} \int_0^\infty
\frac{\theta_{xy}}{\sigma_{\theta}^2}
\left(1-\exp\left(-\dfrac{\theta_{xy}^2}{2\sigma_{\theta}^2}\right)\sum_{m=0}^M \mathcal{H}(m) 
\left(\dfrac{\theta_{xy}^2}{\sigma_{\theta}^2}\right)^m \right)	\nonumber \\
& \times \exp\left[\left({\dfrac{Z^2}{2\sigma_p^2}-\frac{1}{2\sigma_{\theta}^2}}\right)\theta_{xy}^2\right]
Q\left( \left(w_{z_{eq}}^2\ln\left(\dfrac{h}{A_0h_l}\right) + 6Z^2\theta_{xy}^2 + C_b\right)\bigg/\sqrt{32\sigma_p^2Z^2\theta_{xy}^2+C_c}\right)
d\theta_{xy}. 
\end{align} 
\hrulefill
\end{figure*}
%
%
\section{System Model and Main Assumptions for UAV-to-UAV Link}
\label{sec:sysmodel}
We assume that the UAV nodes hovers at the determined locations in the sky.  
For these UAV-based transmission systems, we have three different links: ground-to-UAV (GU), UAV-to-ground (UG) and UAV-to-UAV (UU) links. 
In what follows, we focus on the UU link\footnote{The reader is urged to see \cite{mypaper} for a more in depth channel modeling of UAV-based FSO links and \cite{Dabiri-Savoj-Sadough-WACOWC2019} for the special case of GU link.}.

The probability density function (PDF) of the UU atmospheric channel $h$ is expressed as \eqref{uu} \cite[Eq. (49)]{mypaper}, where 
$C_a = \gamma^2/\left(A_0h_l\right)^{\gamma^2} \times \exp\left(2\sigma_{Lnh_a}^2\gamma^2(1+\gamma^2) \right)$,
$C_b = 2\sigma_{Lnh_a}^2w_{z_{eq}}^2(1+2\gamma^2)$, 
$C_c = 4\sigma_{Lnh_a}^2 w_{z_{eq}}^4$, 
$\gamma^2=w_{z_{eq}}^2/8\sigma_p^2$, with
$ A_0=|\mathrm{erf}(\nu)|^2 $ denoting the maximal fraction of the collected intensity where $ \nu=\frac{\sqrt{\pi}r_a}{\sqrt{2}w_z} $, $w_z$ the optical beam waist at the aperture of receiver, $ w_{z_{\rm eq}}^2=w_z^2\dfrac{\sqrt{\pi}\mathrm{erf}(\nu)}{2\nu\exp(-\nu^2)} $ the equivalent beam width, 
$\delta(.)$ the well-known Dirac delta function, $\mathrm{erfc}(.)=[1-\mathrm{erf}(.)]$ the complementary error function and finally, $Q(.)$ the {\it Q}-function.
In \eqref{uu}, $\mathcal{H}(m)$ is related to the receiver field-of-view angle $\theta_{FOV}$, defined as 
\begin{align}
\label{Hmm}
\mathcal{H}(m) = \frac{2^{-m}M^{(1-2m)}\Gamma\left(m+1,\frac{\theta_{FOV}^2}{2\sigma_\theta^2}\right)   \Gamma(M+m) }
{\Gamma(M-m+1)\left(\Gamma(m+1)\right)^2},   
\end{align}
and $\Gamma(.)$ is the well-known {\it Gamma} function and $\Gamma(.,.)$ denotes the upper incomplete gamma function \cite{jeffrey2007table}. 
Moreover, $h_l$ denotes atmospheric attenuation, $\sigma_{Lnh_a}^2$ denotes the variance of log-irradiance, $\sigma_p^2$ denotes the variance of fluctuations of UAV's position, $\sigma_{\theta}^2$ denotes the variance of deviation of UAV's orientation, and $Z$ is the optical link length.

%
\section{Throughput Analysis of the UU Link}
\label{sec:ergCap}
The throughput of the FSO system is written as \cite{tse2005fundamentals}
\begin{align}
\label{capacity}
\mathcal{C}_{\rm erg}=\int_0^\infty \log\left(1 + \dfrac{R^2 P_t^2 h^2}{\sigma_n^2}\right)f_h(h)dh,
\end{align}
where $P_t$ is the transmitted optical power, $R$ is the responsitivity of photodetector and $\sigma_n^2$ is the variance of unwanted receiver noise $n_o$ and $f_h(h)$ is the probability density function of the instantaneous channel coefficients.
In this work, we suppose that $n_o$ is a Gaussian signal-independent noise with zero-mean and variance $\sigma_{n}^2= 2B_e\, R\, e\, P_b$, where $e$ is the electron charge, $P_b$ is the undesired background power collected by field-of-view of receiver, and $B_e$ is the bandwidth of the receiver's photo detector (in Hz). Based on to \cite{kopeika1970background} and \cite{mypaper}, undesired background power is a function of receiver's field-of-view which is denoted by $\theta_{FoV}$ and $P_b=\frac{\pi}{4} B_o\, N_b(\lambda)\,  A_a\,\theta_{FoV}^2$, where $N_b(\lambda)$ is the spectral radiance of the background radiations at wavelength $\lambda$ (in Watts/${\rm cm}^2$-$\micro$m-srad),
$A_a$ is the area of receiver lens (in ${\rm cm}^2$),
and $B_o$ denotes the optical bandwidth (mainly limited by optical filter) at the receiver (in $\micro$m).

For higher values of signal-to-noise ratio (SNR), the ergodic capacity or ergodic throughput expressed in \eqref{capacity} can be approximated as \cite{tse2005fundamentals}
\begin{align}
\label{lower}
\mathcal{C}_{\rm erg} \simeq \int_0^\infty \log\left(\dfrac{R^2 P_t^2 h^2}{\sigma_n^2}\right)f_h(h)dh. 
\end{align}
By substituting $f_h(h)$ in \eqref{lower} by $f^{uu}_h(h)$ from \eqref{uu}, we get
\begin{align}
\label{d2n1}
\mathcal{C}_{\rm erg}^{uu}\simeq &~  	
C_a \int_0^{\infty} \int_0^\infty    
\ln\left(\dfrac{R^2 P_t^2 h^2}{\sigma_n^2}\right)   h^{\gamma^2-1} \frac{\theta_{xy}}{\sigma_{\theta}^2}  \nonumber \\
&\times 
\left(1-\exp\left(-\dfrac{\theta_{xy}^2}{2\sigma_{\theta}^2}\right)\sum_{m=0}^M \mathcal{H}(m) 
\left(\dfrac{\theta_{xy}^2}{\sigma_{\theta}^2}\right)^m \right)	\nonumber \\
&\times Q\left( \dfrac{w_{z_{eq}}^2\ln\left(\dfrac{h}{A_0h_l}\right) + 6Z^2\theta_{xy}^2 + C_b}{\sqrt{32\sigma_p^2Z^2\theta_{xy}^2+C_c}}\right) \nonumber \\
& \times \exp\left[\left({\dfrac{Z^2}{2\sigma_p^2}-\frac{1}{2\sigma_{\theta}^2}}\right)\theta_{xy}^2\right] 
d\theta_{xy} dh. 
\end{align} 
On the other hand, the {\it Q}-function in \eqref{d2n1} can be simplified as \cite{zhang2014block}
\begin{align}
\label{q_func}
Q(x) \simeq \sum_{i=1}^3 a_i\times e^{(-a'_ix^2)},
\end{align}
where $a_1=\frac{5}{24}$, $a_2=\frac{4}{24}$, $a_3=\frac{1}{24}$, $a'_1=2$, $a'_2=\frac{11}{20}$, and $a_3=\frac{1}{2}$,.
By using \eqref{q_func} to simplify the term $Q\left( \dfrac{w_{z_{eq}}^2\ln\left(\dfrac{h}{A_0h_l}\right) + 6Z^2\theta_{xy}^2 + C_b}{\sqrt{32\sigma_p^2Z^2\theta_{xy}^2+C_c}}\right)$ in \eqref{d2n1} and after using a change of variable $\hbar=\ln (h)$, after some algebra, we have
\begin{align}
\label{sb41}
\mathcal{C}_{\rm erg}^{uu} \simeq&~ C_a \sum_{i=1}^3 a_i \int_0^\infty \int_{-\infty}^{\infty}
\frac{\theta_{xy}}{\sigma_{\theta}^2}
\exp\left[\left({\dfrac{Z^2}{2\sigma_p^2}-\frac{1}{2\sigma_{\theta}^2}}\right)\theta_{xy}^2\right]	\nonumber \\
%
&\times  \left(K_1+2\hbar\right)  \exp\left( \dfrac{J_{2i}(\theta_{xy})\gamma^4}{2} -J_1(\theta_{xy})\gamma^2 \right) 
\nonumber \\
&\times \left(1-\exp\left(-\dfrac{\theta_{xy}^2}{2\sigma_{\theta}^2}\right)\sum_{m=0}^M \mathcal{H}(m) 
\left(\dfrac{\theta_{xy}^2}{\sigma_{\theta}^2}\right)^m \right) \nonumber \\
%
&\times \exp\left( -\dfrac{\left(\hbar +J_1(\theta_{xy})-J_{2i}(\theta_{xy})\gamma^2\right)^2 }{{2J_{2i}(\theta_{xy})}} \right)
d\hbar d\theta_{xy},  
\end{align} 
%
where 
$J_1(\theta_{xy})=\dfrac{6Z^2\theta_{xy}^2 + C_b}{w_{z_{eq}}^2}- \ln(A_0h_l)$,
$J_{2i}(\theta_{xy})=\dfrac{16\sigma_p^2Z^2\theta_{xy}^2+C_c}{a_i' w_{z_{eq}}^4}$ and 
$K_1 = \ln\left(\frac{R^2 P_t^2}{\sigma_n^2}\right)$.
According to \cite{jeffrey2007table} and after some mathematical calculations, the two-dimensional integration of \eqref{sb41} is reduced to a one-dimensional integration as 
\begin{align}
\label{uu_4}
\mathcal{C}_{\rm erg}^{uu} \simeq   \underbrace{(\mathbb{I}_{11} +  \mathbb{I}_{12})}_{\mathbb{I}_1}
-   \underbrace{(\mathbb{I}_{21} + \mathbb{I}_{22})}_{\mathbb{I}_{2}},
\end{align}
where $\mathbb{I}_{11}$, $\mathbb{I}_{12}$, $\mathbb{I}_{21}$ and $\mathbb{I}_{22}$ are derived in \eqref{I_11}, \eqref{I_12}, \eqref{I_21} and \eqref{I_22}, respectively. In \eqref{I_11}-\eqref{I_22}, we have 
\begin{align}
&K_{2i} = C_a(A_0h_l)^{\gamma^2}\!\sqrt{\frac{32\pi a_i^2  \sigma_p^2Z^2}{a_i'\sigma_{\theta}^4  w_{z_{eq}}^4}}
\!\exp\left( \!  \frac{C_c-16a_i'\sigma_p^2C_b}{128 a_i'\sigma_p^4}  \! \right)\!, \nonumber \\
& K_{3i} = \frac{C_c-8 a_i' \sigma_p^2 C_b}{4 a_i' \sigma_p^2 w_{z_{eq}}^2}    +2\ln(A_0h_l) + K_1, \nonumber \\
& K_{4i} = \left(\frac{(6a_i'-1) Z^2}{8 a_i' \sigma_p^2} + \frac{1}{2\sigma_{\theta}^2} -  \dfrac{Z^2}{2\sigma_p^2}\right). \nonumber
\end{align}
%
%
\begin{figure*}
\normalsize
\begin{align}
\label{I_11}
&\mathbb{I}_{11}=\sum_{i=1}^3  K_{2i}
\frac{(4-12 a_i')Z^2}{a_i'  w_{z_{eq}}^2}  	
\int_0^\infty   
\sqrt{\theta_{xy}^8 +\frac{C_c\theta_{xy}^6}{16\sigma_p^2Z^2}} 	
\exp\left(-K_{4i}\theta_{xy}^2\right)  
d\theta_{xy},   \\
\label{I_12}
&\mathbb{I}_{12}=\sum_{i=1}^3  K_{2i} K_{3i}
\int_0^\infty 
\sqrt{\theta_{xy}^4 +\frac{C_c\theta_{xy}^2}{16\sigma_p^2Z^2}} 	
\exp\left(-K_{4i}\theta_{xy}^2\right)  
d\theta_{xy},   \\
\label{I_21}
& \mathbb{I}_{21} = \sum_{i=1}^3  K_{2i}
\frac{(4-12 a_i')Z^2}{a_i'  w_{z_{eq}}^2}   
\int_0^\infty 
\sqrt{\theta_{xy}^6 +\frac{C_c\theta_{xy}^4}{16\sigma_p^2Z^2}} 
\sum_{m=0}^M \mathcal{H}(m) 
\left(\dfrac{\theta_{xy}}{\sigma_{\theta}}\right)^{2m+1} 	
\exp\left(-K_{4i}'\theta_{xy}^2\right)  
d\theta_{xy},  \\
\label{I_22}
& \mathbb{I}_{22} = \sum_{i=1}^3  K_{2i} K_{3i}
\int_0^\infty 
\sqrt{\theta_{xy}^2 +\frac{C_c}{16\sigma_p^2Z^2}} 
\sum_{m=0}^M \mathcal{H}(m) 
\left(\dfrac{\theta_{xy}}{\sigma_{\theta}}\right)^{2m+1} 	
\exp\left(-K_{4i}'\theta_{xy}^2\right)  
d\theta_{xy}.  
\end{align} 
\hrulefill
\end{figure*}
Applying a change of variable $\theta_{xy}^2 +\frac{C_c}{16\sigma_p^2Z^2}=x_{11}$ and using \cite[Eq. (3.383.4)]{jeffrey2007table} and after some mathematical derivations, the closed-form analytical expression for $\mathbb{I}_{11}$ is obtained as 
\begin{align}
\label{i_11}
&\mathbb{I}_{11} 
= 1.722\sum_{i=1}^3 K_{2i} K_{4i}^{-7/4}  \frac{(1-3 a_i')Z^2}{a_i'  w_{z_{eq}}^2}
\exp\left(\frac{K_{4i} C_c}{16\sigma_p^2Z^2} \right)  \nonumber \\
&\times \left(\frac{C_c}{16\sigma_p^2Z^2}\right)^{3/4}  
\exp\left(- \frac{K_{4i}C_c}{32\sigma_p^2Z^2}\right)
W_{-\frac{1}{4}, -\frac{5}{4}}\left( \frac{K_{4i}C_c}{16\sigma_p^2Z^2}\right),
\end{align} 
where $W_{a,b}\left( . \right)$ is the Whittaker function \cite{wolfram1} that is a modified form of the confluent hypergeometric equation.
Also, by applying a change of variable $\theta_{xy}^2=x_{12}$ and using \cite[Eq. (3.382.4)]{jeffrey2007table}, the closed-form expression for $\mathbb{I}_{12}$ is derived as
\begin{align}
\label{sb60}
\mathbb{I}_{12} =&  \sum_{i=1}^3 \frac{ K_{2i} K_{3i} K_{4i}^{-3/2} }{2}   \exp\left( \frac{K_{4i} C_c}{16\sigma_p^2Z^2}   \right) 
\Gamma\left(\frac{3}{2},  \frac{K_{4i} C_c}{16\sigma_p^2Z^2}   \right), 
\end{align} 
where $\Gamma(.,.)$ is the incomplete  Gamma  function \cite{wolfram1}.
Similar to the derivation of \eqref{i_11}, the closed-form expressions for $\mathbb{I}_{21}$ and $\mathbb{I}_{22}$ are derived respectively as
\begin{align}
\label{i_21}
\mathbb{I}_{21} =& \sum_{i=1}^3 \sum_{m=0}^M    \Gamma(m\!+\!2)  \frac{\mathcal{H}(m) K_{2i}  K_{4i}^{-\frac{2m+7}{4}} }{ 2 \sigma_{\theta}^{2m} }  
\frac{(4-12 a_i')Z^2}{a_i'  w_{z_{eq}}^2} \nonumber \\
&\times \exp\left(\frac{K_{4i} C_c}{32\sigma_p^2Z^2} \right) 
\left( \frac{C_c}{16\sigma_p^2Z^2}  \right)^{\frac{2m+3}{4}} \nonumber \\
&\times W_{-\frac{2m+1}{4},-\frac{2m+5}{4}}\left(\frac{K_{4i} C_c}{16\sigma_p^2Z^2} \right),
\end{align} 
and
\begin{align}
\label{i_22}
\mathbb{I}_{22}=& \sum_{i=1}^3 \sum_{m=0}^M   \Gamma(m+1)    \frac{\mathcal{H}(m) K_{2i} K_{3i} K_{4i}^{-\frac{2m+5}{4}} }{ 2 \sigma_{\theta}^{2m} }  \nonumber \\ 
&\times \exp\left(\frac{K_{4i} C_c}{32\sigma_p^2Z^2} \right)  
\left( \frac{C_c}{16\sigma_p^2Z^2}  \right)^{\frac{2m+1}{4}} \nonumber \\  
&\times W_{-\frac{2m-1}{4},-\frac{2m+3}{4}}\left(\frac{K_{4i} C_c}{16\sigma_p^2Z^2} \right).
\end{align} 
Finally, by substituting \eqref{i_11}-\eqref{i_22} in \eqref{uu_4}, the closed-form expression for $\mathcal{C}_{\rm erg}^{uu}$ is obtained.

According to \eqref{Hmm}, $H(m)$ depends solely on $G=\frac{\theta_{FOV}}{\sigma_\theta}$, where $G$ increases by increasing $\theta_{FOV}$. On the other hand, $H(m)$ tends to zero by increasing $G$. Hence, for higher values of $G$ which is obtained for higher values of $\theta_{FOV}$, we can neglect the therm $\mathbb{I}_2$ compared to $\mathbb{I}_1$ and according to \eqref{sb41} and \eqref{uu_4}, we get 
\begin{align}
\label{nb2}
\mathcal{C}_{\rm erg}^{uu} \simeq \mathbb{I}_1.
\end{align} 
\section{Numerical Analysis}
\label{sec:simul}
In the sequel, we provide numerical and simulation results to study the performance of optical UAV-based FSO communication systems in terms of achievable rate (mean ergodic capacity).
Simulation results are performed to investigate the accuracy of the derived closed-form expressions.
Our aim is to investigating the impact of receiver and transmitter deviations, beam width $w_z$, optical link length $Z$, and field-of-view angle $\theta_{FOV}$ on the performance of UU and UG communications.
The value of parameter systems used for performance analysis is provided in Table I.
%
\begin{table}
\caption{Nomenclature and Simulation Setup} 
\centering 
\begin{tabular}{l c} 
	\hline\hline \\[-1.2ex]
	{\bf Parameter (Symbol)} & {\bf Numerical Value \& Unit} \\ [.5ex] 
	\hline\hline \\[-1.2ex]
	Wavelength $(\lambda)$            & 1550 nm   \\[1ex] 
	Aperture Radius $ (r_a) $           &  5 cm \\[1ex] 
	Responsivity of PD $(R)$                  & 0.6 \\[1ex]                                                                                   
	Bandwidth of PD $ (B_e) $           & 1 GHz  \\[1ex]  
	Bandwidth of optical filter $ (B_o) $           & 10 nm \\[1ex]
	Spectral radiance $ (N_b(\lambda)) $   & $10^{-3}$ W/${\rm cm}^2$-µm-srad \\[1ex]
	variance of log-irradiance $(\sigma^2_{{\rm Ln}h_a})$     & 0.1 \\[1ex]
	Atmospheric Attenuation $(h_l)$     & 0.1 \\[1ex]
	SD of position fluctuations $ (\sigma_p) $                   & 25 cm\\[1ex]         
	\hline \\[-2ex]
	Link length $(Z)$         & 200 m \\[1ex]
	SD of UAV's orientation deviations $(\sigma_{\theta})$ & 1-12 mrad \\[1ex] 
	FOV angle $(\theta_{FOV})$  &  1-40 mrad \\[1ex]
	Beam width at Rx $(w_z)$ & 0.2-6 m \\[1ex]
	\hline\hline 
\end{tabular}
\label{I} 
\end{table}
%

%
\begin{figure}[!h]
\begin{center}
	\includegraphics[width=3.3 in ]{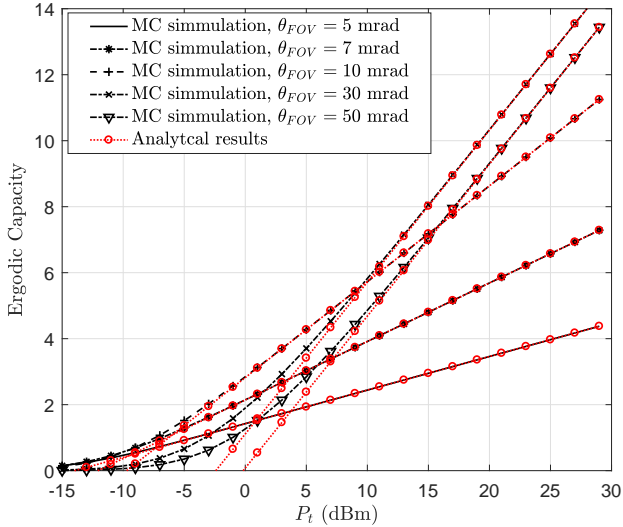}
	\caption{Ergodic capacity of UU link versus $P_t$ for different values of $\theta_{FOV}$ and $w_z=2$ m and $\sigma_\theta=5$ mrad.  }
	\label{Sn1}
\end{center}
\end{figure}
%

First, we address the effect of receiver FOV which is a main tunable parameter impacting the performance of the considered communication system.
To this end, we have depicted in Fig. \ref{Sn1} the achievable rate or ergodic capacity of UU optical link versus a wide range of $P_t$ for different values of receiver FOV angle. The receiver beam width and SD of UAV orientation deviations are fixed at $w_z=2$ m and $\sigma_{\theta}=5$ mrad. As we observe, at high SNR, we see a good match between analytical and simulation results. The main insight that can be drawn from Fig. \ref{Sn1} is that by changing the field of view angle, the capacity of optical UAV-based link is altered significantly.


Results shown in Fig. \ref{Sn1} indicate that 
the capacity of a UAV-based optical wireless communication systems system highly depends on the optimal values for tunable parameters $\theta_{FOV}$ and $w_z$ and this confirms the importance of selecting the optimal values for considered adjustable parameters.
This raises the the following fundamental question: for a given channel condition (i.e., for a given $\sigma_{\theta}$, $P_t$, for instance), what are the maximal achievable data rates and how to select the tunable parameters to maximize the ergodic capacity ?
\begin{figure}
\centering
\subfloat[] {\includegraphics[width=3.3 in]{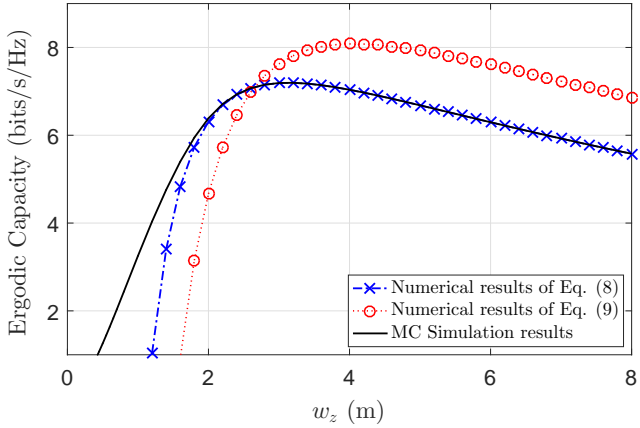}
	\label{M1}
}
\hfill
\subfloat[] {\includegraphics[width=3.3 in]{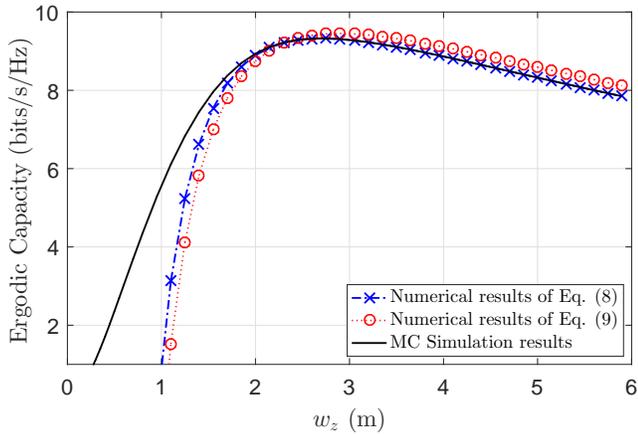}
	\label{M2}
}
\hfill
\subfloat[] {\includegraphics[width=3.3 in]{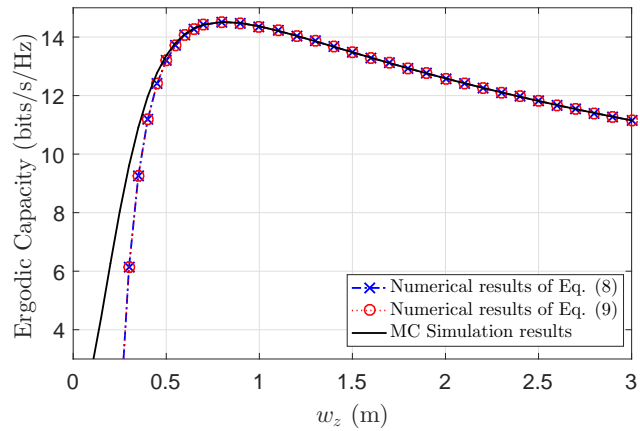}
	\label{M3}
}
\caption{Ergodic capacity of UU link versus $w_z$ for $P_t=10$ dBm, $\theta_{FOV}=25$ mrad; a) $\sigma_\theta=10$, b) $\sigma_\theta=7$, c) $\sigma_\theta=2$ mrad.}
\label{M4}
\end{figure}
%

To get insights about the effect of beam width on ergodic capacity, in the sequel, we focus on the behavior of the ergodic capacity of optical UAV-based communication systems versus beam width $w_z$. The optimum value of $w_z$ refers to the value of $w_z$ for which the maximum capacity is achieved.
In Fig. \ref{M4}, we have depicted the capacity associated to the UU communication link as a function of $w_z$  for $\sigma_{\theta}=$ 10, 7 and 2 mrad.
As can be seen, the optimum values of $w_z$  increases  by increasing $\sigma_{\theta}$.
From Fig. \ref{M4}, we show a high correlation between the average capacity and $w_z$. This is specially suitable to design UAV-based communication systems according to the worst instantaneous channel conditions (i.e., the channel with higher orientation and position instability).
In this way, when $\sigma_{\theta}$ decreases, we will be sure that the throughput remains approximately near its maximum value. To clarify more this issue, consider for instance a channel with $\sigma_{\theta}$ changing in the interval $[2, 10]$ mrad due to the variation in the wind speed during a day. According to results depicted in Fig. \ref{M1}, in the worst case characterized by $\sigma_{\theta}=10$ mrad, the optimum value for $w_z$ is 3 m. Now, according to the results of Fig. \ref{M3}, if $w_z=3$ is used for the best case i.e., when $\sigma_{\theta}=2$ mrad, the achievable rates is 11.2 {\rm bits/s/Hz} which is only 3.3 {\rm bits/s/Hz} lower than the maximum possible achievable rate. 

According to the results presented in Figs. \ref{M1}-\ref{M3}, by decreasing $\sigma_{\theta}$ that leads to an increase in $G$, the ergodic capacity \eqref{nb2} becomes close to that derived in \eqref{uu_4}. 
For instance, we observe from Fig. \ref{M2} that for $G=3.6$, the ergodic capacity obtained with \eqref{nb2} is close to that derived with \eqref{uu_4}. Notice that the gap between the simulation and analytical results at low values of $w_z$ is due to the approximation used in \eqref{lower}. However, by increasing $w_z$ this gap vanishes, especially for optimum values of $w_z$.

\section{Concluding Remarks}
\label{sec:concl}
The problem of achievable data rates in UAV-based FSO communications was investigated. 
We showed that in such systems, in addition to the goemetrical pointing error due to the UAV's position deviations and instantaneous atmospheric turbulence coefficients, angle of arrival fluctuations due to orientation fluctuations of hovering UAVs is a vital factor which limits the achievable data rates.
Based on a characterization of the aforementioned system parameters, we derived the exact analytical expressions for average capacity of UAV-to-UAV FSO  communication link.  Then, we studied the impact of main parameters of FSO system on average capacity of the considered system. 
It was shown that in addition to the orientation fluctuations of the receiver, which affect angle of arrival deviations, orientation deviations of the transmitter node cause deviations in the center of receiver's lens and received optical beam foot print and has a more important destructive effect on the capacity of UAV based FSO links. Moreover, the ergodic capacity of the considered system was shown to be strongly dependent on the tunable parameters $w_z$ and $\theta_{FOV}$, 
where the optimal amount of aforementioned parameters varies by changing the strength of orientation fluctuations of drone. Monte Carlo simulations verified a good match between simulation and analytical results especially for larger values of signal to noise ratio. The analytical methods can be used in parameter optimization for maximizing the average capacity of UAV-based optical wireless communication systems without employing to cumbersome  Monte-Carlo simulations. 


\end{document}